# Timed k-Tail: Automatic Inference of Timed Automata

Fabrizio Pastore, Daniela Micucci, Leonardo Mariani
Department of Informatics, Systems, and COmmunications
University of Milano - Bicocca
Milano, 20126
Email: {pastore,micucci,mariani}@disco.unimib.it

***Abstract*—Accurate and up-to-date models describing the behavior of software systems are seldom available in practice. To address this issue, software engineers may use specification mining techniques, which can automatically derive models that capture the behavior of the system under analysis.**

**So far, most specification mining techniques focused on the functional behavior of the systems, with specific emphasis on models that represent the ordering of operations, such as temporal rules and finite state models. Although useful, these models are inherently partial. For instance, they miss the timing behavior, which is extremely relevant for many classes of systems and components, such as shared libraries and user-driven applications. Mining specifications that include both the functional and the timing aspects can improve the applicability of many testing and analysis solutions.**

**This paper addresses this challenge by presenting the Timed k-Tail (*TkT*) specification mining technique that can mine timed automata from program traces. Since timed automata can effectively represent the interplay between the functional and the timing behavior of a system, *TkT* could be exploited in those contexts where time-related information is relevant.**

**Our empirical evaluation shows that *TkT* can efficiently and effectively mine accurate models. The mined models have been used to identify executions with anomalous timing. The evaluation shows that most of the anomalous executions have been correctly identified while producing few false positives.**

## I. Introduction

Behavioral descriptions are extremely useful to effectively test and analyze software systems. For instance, behavioral information can be used to generate test cases [1], [2], identify anomalies [3], [4], detect inefficiencies [5], and support debugging [6], [7], [8]. Unfortunately, suitable behavioral descriptions are seldom available, and in many cases the only source of information about the behavior of a system is the system itself.

So far, specification mining techniques have been extensively used to inexpensively produce specifications that can be exploited to support software engineering tasks [1], [9], [10], [11], [12], [13]. In particular, specification mining can distill behavioral models that precisely represent the behavior of the monitored system by observing its runtime behavior. Specification mining techniques can be used to generate a variety of models about the *functional* behavior of a system, including finite state models [10], [11], [13], extended finite state machines [14], [15], temporal rules [16], [17], and invariants [12]. While functional aspects have been extensively covered by specification mining techniques, little attention has been paid to the *interplay between the functional and the timing aspects*. The interplay between these two aspects is relevant to understand the behavior of a system. For instance, functionally correct behaviors might be wrong when looking at the timing aspect, such as the case of an operation that produces the correct result but takes too much time to complete. Vice versa behaviors that are correct according to the timing aspect might be wrong when looking at the functional aspect, such as the case of an operation that completes quickly by producing a wrong result.

So far, the problem of mining specifications that show how the functional behavior of a program relates to its timing behavior has been addressed with techniques that derive timed automata [18], [19], [20], [21], [22], [23], models that represents both functional and timing information. Unfortunately, most of these techniques are specific to systems that process signals and control actuators, and can be hardly applied to other software systems [18], [20]. In addition, the most frequently inferred model is the *one clock* timed automaton [24] that cannot be used to represent execution time boundaries in presence of nested operations, which are common in many software systems.

A specification mining technique that can be applied to a broader set of systems is Perfume [5], which can produce a finite state model annotated with information about resource consumption (e.g., time consumption). Perfume is not meant to be a pure specification mining solution, rather it has been designed to produce models that can isolate the behaviors with a suspicious resource consumption. This capability is assumed to facilitate the identification of the suspicious behaviors, but may also negatively affect the overall capability to accurately and efficiently represent the behavior of the observed system, as observed in our experiments.

To address the challenge of mining accurate models that cover both the functional and timing aspects, we designed *Timed K-Tail* (*TkT*), a specification mining technique that can target any kind of software system whose runtime behavior can be represented as the execution of a possibly nested sequence of operations (e.g., sequences of method calls). In particular, *TkT* starts from a set of traces with information about the operations that have been executed and their durations, and mines a *Timed Automaton* that captures the general behavior of the software. To achieve this result, *TkT* extends the k-Tail automaton learning technique [25] with the capability to generate clocks, reset operations, and guard conditions.

In our empirical evaluation based on traces collected from various algorithms and libraries, we show that *TkT* can be both more effective and efficient than Perfume.

The major contributions of this paper are:

- The definition of *TkT*, a specification mining technique for the generation of timed automata that can be applied to any software system whose behavior can be represented as possibly nested sequences of operations,
- The implementation of a publicly available version of *TkT*, downloadable from http://www.lta.disco.unimib.it/tools/tkt/,
- An empirical evaluation that compares *TkT* to the Perfume state of the art technique.

The paper is organized as follows. Section II presents the *TkT* specification mining technique using a simple running example. Section III presents our empirical evaluation. Section IV discusses related work. Section V provides final remarks.

## II. Timed k-Tail

Given a set of traces collected from a running system, Timed K-Tail (*TkT*) can automatically generate a timed automaton that generalizes the observations in the traces. The generated timed automaton represents the behavior of the monitored system in terms of both the sequences of operations that the system can execute and the timing of these operations.

*TkT* extends the k-Tail algorithm [25], which can only produce simple automata, with the ability to generate clocks, reset operations, and clock constraints. In the following, we present the *TkT* algorithm with specific emphasis on the extensions that we introduced with respect to the k-Tail algorithm.

### A. Overview

Figure 1 gives a visual overview of the inputs and outputs of the five steps of the *TkT* algorithm. The input to *TkT* is a set of timed traces with information about the ordering and the timing of the operations executed by a program, see for instance *Trace 1* and *Trace 2* in Figure 1. For each executed operation, a trace includes two events that mark the beginning and the end of the operation, respectively. We graphically indicate the beginning and the end of a same operation with the letters *B* and *E* respectively, plus a thick line that connects them. For instance, the first two events in *Trace 1* represent the beginning and the end of the `ProcessWebOrder` operation. Finally, each event has a timestamp indicating when the event has been observed.

In the first step, *Trace Normalization*, *TkT* normalizes the timing information in the traces, which might have been collected at different times, by assigning time 0 to the first event in each trace and adjusting the times associated with the other events in the traces consistently to preserve time difference between events. For instance, the time associated with the first two events of *Trace 1* is changed from 98483940 and 98483943 to 0 and 3, respectively. This operation is important to align time across multiple traces.

In the second step, *Automaton Initialization*, *TkT* produces an initial automaton where each trace corresponds to a branch in the model, see for instance the *Initial Automaton* in Figure 1. The automaton is annotated with *relative clocks* that measure the time taken by each operation logged in the trace to complete. In particular, each clock is reset when an operation starts and checked with an equality constraint when the operation completes. Each instance of a (same) operation is associated with a different relative clock. For instance, the duration of the first occurrence of the `processItem` operation in the *Normalized Trace 1* is measured by clock `c2` in the automaton, while the second occurrence of the same operation in the same normalized trace is measured by clock `c4`. The automaton also defines an *absolute clock* `t` that is reset when the first operation is executed and checked at every transition of the model with an equality constraint.

In the third step, *State Merging*, *TkT* merges equivalent states, which are states in the model that are likely to represent a same state of the monitored system. Merging states might generate transitions that start and end from the same states and have the same label, that is redundant transitions that represent a same event. In such a case, *TkT* merges these transitions into a single transition with the same label and merged annotations (i.e., the resulting transition includes all the reset and clock constraints in the merged transitions). Thus the more states are merged the more reset operations and equality constraints are accumulated on the same transitions of the model. After the state merging process has completed, the resulting model represents the behavior of the monitored software quite extensively, see for instance the *Merged Automaton* in Figure 1. Note that while the information about the sequences of operations that can be performed is a generalization of the sequences reported in the traces, the timing information is still encoded as a simple set of equality constraints that match the observations reported in the traces. The next two steps of the process further elaborate the timing information to produce more general and flexible timing constraints.

In the fourth step, *Clock Refinement*, *TkT* identifies the occurrences of distinct clocks that measure the duration of exactly the same operations, that is redundant clocks that are reset and checked on the same transitions, and transforms them in a same clock. This transformation reduces the number of clocks in the model and increases the number of equality constraints on a same clock associated with a same transition. See for instance the *Automaton with Refined Clocks* in Figure 1.

In the last step, *Guards Generation*, *TkT* processes the information on each transition to transform the many equality constraints on the individual clocks into a time interval constraint. If only one sample is available for a clock, the equality constraints are discarded and no interval is generated. In this phase, the user may decide to incorporate or discard the constraints on the absolute time, depending on the confidence over the robustness of such constraints. The resulting model is a timed automaton that captures not only the ordering of the events, but also the expected duration of the operations. See for instance the *Timed Automaton* in Figure 1.

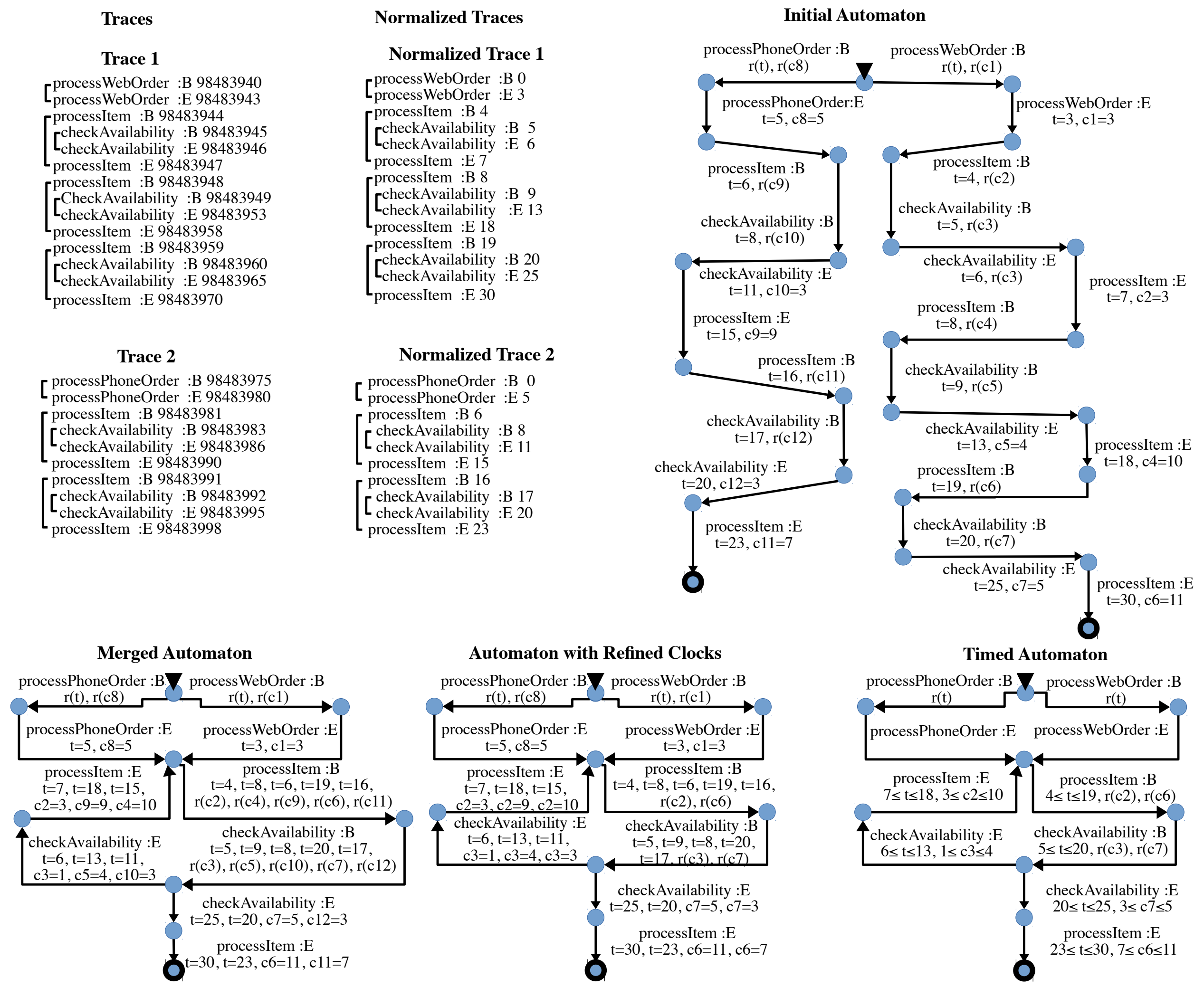

Fig. 1. Running Example: TKT inputs, intermediate results, and output.

In the following, we detail each step of the algorithm.

### B. Trace Normalization

*TkT* processes timed traces, that is traces that log the execution of the operations and their durations. Formally a *timed trace* is defined as a temporally ordered sequence $event_1 \dots event_n$, where $event_i = (type_i, op_i, time_i)$, with

- $type \in \{\uparrow, \downarrow\}$, where $\uparrow$ and $\downarrow$ indicate the beginning and the end of the execution of an operation, respectively,
- *op* is a label that identifies the operation that has started or ended,
- $time \in \mathbb{N}_0^+$ is the timestamp of the event.

Given an event $e$, we denote with $type(e)$, $op(e)$, and $time(e)$ the corresponding element of $e$.

A well-formed timed trace satisfies the following properties:

- *time does not decrease*: $time_i \leq time_j \; \forall i \leq j$
- *events are paired*: when an operation *op* starts, it produces the event $(\uparrow, op, time_i)$ and when it ends, it produces the event $(\downarrow, op, time_j)$ with $time_i \leq time_j$. Given an event $e$ of type either $\uparrow$ or $\downarrow$ referring to an operation *op*, we indicate the event of the other type referring to the same operation *op* with $pair(e)$.
- *nesting of operations is respected*: all the operations started after the beginning of an operation *op* must have ended before the operation *op* ends.

Each trace starts with its own timestamp. This is not an issue for relative clocks, but when the inferred timed automaton includes constraints on the absolute time, it is important to make sure all the traces refer to the same starting time, otherwise the timing information would not be comparable.

We conventionally use 0 as starting time of each trace, thus the normalization process simply subtracts to the timestamp of each event the value of the timestamp of the first event in the trace.

### C. Automaton Initialization

The *Automaton Initialization* step generates an *Initial Automaton* that accepts all and only the executions stored in the traces (see Figure 1). The *Initial Automaton* is obtained by mapping each trace to an independent branch of the automaton. Here we first report a formal definition of timed automaton and then formally present how *TkT* generates the *Initial Automaton* from the normalized traces.

A *Timed Automaton* is a tuple $(S, s_0, C, E, \mathit{TR})$, where

- $S$ is a finite set of states,
- $s_0 \in S$ is the initial state,
- $C$ is a finite set of clocks,
- $E$ is a finite set of events,
- $\mathit{TR} \subseteq (S \times E \times \{\uparrow,\downarrow\} \times G(C) \times 2^C \times S)$ indicates the set of transitions $tr = (s_a, e, t, g, r, s_b)$ from a state $s_a$ to state $s_b$ annotated with an event label $e$, an event type $t$ (equals to either $\uparrow$ or $\downarrow$), a set of guards $g$ on the set of clocks $C$, and a set $r$ of clocks reset to 0.

We do not include state invariants in our definition because *TkT* only generates guards on transitions.

Given a timed trace $tt = \mathit{event}_1 \ldots \mathit{event}_n$, with $\mathit{event}_i = (\mathit{type}_i, \mathit{op}_i, \mathit{time}_i)$ and a timed automaton $TA = (S, s_0, C, E, \mathit{TR})$, TA *accepts tt* if and only if $\exists tr_1 \ldots tr_n$, with $tr_i = (s_{i-1}, e_i, t_i, g_i, r_i, s_i) \in TR\ \forall i = 1 \ldots n$ that satisfies the following conditions: $e_i = \mathit{op}_i$, $t_i = \mathit{type}_i$, and each constraint in $g_i$ evaluates to true, according to the values of the clocks. Note that the value of a clock is determined by the time passed between the last reset operation executed on that clock and the timestamp of the current event in the trace. Guards on clocks that have never been reset are ignored.

Given a set of timed traces $T$, the *Automaton Initialization* step creates a timed automaton defined as follows:

- $S = \{s_0\} \cup \{$a state $s_{i,j}$ for each $\mathit{event}_i$ in $\mathit{trace}_j\}$.
- $C$=$\{$the absolute clock $t\} \cup \{$a relative clock $c_{i,j}$ for each $\mathit{event}_i$ of type $\uparrow$ in any $\mathit{trace}_j\}$. Given an event $e$ of type $\uparrow$, we indicate with $\mathit{clock}(e)$ its relative clock.
- $E = \bigcup_{\mathit{trace} \in T} \bigcup_{(\mathit{type},\mathit{op},\mathit{time}) \in \mathit{trace}} \mathit{op}$
- for each $\mathit{event}_i = (\mathit{type}_i, \mathit{op}_i, \mathit{time}_i)$ occurring in a trace $\mathit{trace}_j \in T$, *TkT* adds a transition $tr$ to $TR$ defined as follows $tr = (s_{i-1,j}, \mathit{op}_i, \mathit{type}_i, g, r, s_{i,j})$, where
  - $s_{0,j}$ is the initial state $s_0$ for any value of $j$,
  - $g$ consists of two equalities $t = \mathit{time}_i$, and $\mathit{clock}(\mathit{pair}(\mathit{event}_i)) = \mathit{time}_i - \mathit{time}(\mathit{pair}(\mathit{event}_i))$. The second equality constraints the value of the clock associated to $\mathit{event}_i$ and is present only if $\mathit{type}_i = \downarrow$.
  - if $\mathit{type}_i = \downarrow$, $r = \varnothing$ (clocks are reset only when operations start); otherwise if $i > 1$ then $r = \{c_{i,j}\}$, else $r = \{t, c_{1,j}\}$ (the clock measuring absolute time is reset on the first transition).

### D. State Merging

The *State Merging* step consists of the iterative merging of all the states that accept the same sequences of events. This is a standard strategy introduced in the k-Tail algorithm [25], and also exploited by several other algorithms [5], [13], [14], [26], to produce models that generalize a set of observations. *TkT* defines a version of the state merging process that can handle timing information in addition to sequences of events.

The state merging process starts with the computation of the kFuture of each state, that is the set of all the event sequences of maximum length $k$ that can be accepted by a given state. Since *TkT* takes the type of event (i.e., whether an event indicates either the beginning or the end of an operation) into consideration, the kFuture includes this information.

With $k$ equals to 2, the kFuture of the initial state of the *Initial Automaton* shown in Figure 1 is composed of two sequences of length 2. The first sequence consists of the beginning followed by the end of the execution of the `processWebOrder` operation. The second sequence consists of the beginning followed by the end of the execution of the `processPhoneOrder` operation.

When two states have the same kFuture, they are assumed to represent a same state of the software system, and are thus merged into a single state. This process may change the source and the target states of several transitions, and consequently there might be created redundant transitions that can be merged into a single transition. Two transitions are *redundant* if they start from the same state, end into the same state, and have the same event and type. More formally, given two transitions $tr = (s_a, \mathit{event}, \mathit{type}, g, r, s_b)$ and $tr' = (s'_a, \mathit{event}', \mathit{type}', g', r', s'_b)$, they are redundant if $s_a = s'_a$, $s_b = s'_b$, $\mathit{event} = \mathit{event}'$ and $\mathit{type} = \mathit{type}'$. The merge process drops both transitions and adds a single transition annotated with the values from the two dropped transitions. More formally, the transition added after dropping $tr$ and $tr'$ is $tr_{merged} = (s_a, \mathit{event}, \mathit{type}, g \cup g', r \cup r', s_b)$, essentially the resulting transition accumulates the guard conditions and the reset operations present in the merged transitions.

We call the automaton resulting from this step *Merged Automaton*. Figure 1 shows the *Merged Automaton* obtained from the *Initial Automaton* by applying the state merging process with $k = 2$.

### E. Clock Refinement

The *Merged Automaton* may accept more behaviors than the *Initial Automaton*. However, up to this point, the generalization has only targeted the sequences of events that can be accepted by the automaton, while the timing information still overfits the information in the input traces. For instance, the guard conditions still consist of equality constraints on relative and absolute clocks. It is thus important to generalize the time observations to allow a proper degree of flexibility, so that the model can be used to discover behavioral anomalies on the timing aspect, without being too sensitive to noise.

To soundly estimate the time that might be taken by an operation to complete, it is important to exploit as many observations as possible. Since *TkT* creates a relative clock for each occurrence of each operation in the traces, there is exactly one observation for each relative clock in the model, which is not enough to distill any general information about the timing of the operations. However, the state merging step produces *redundant clocks* that measure the duration of exactly the same behaviors. For instance, clocks $c2$, $c4$, $c9$ in the *Merged Automaton* in Figure 1 all measure the duration of the $processItem$ operation. These individual observations over different relative clocks can be simplified into multiple observations of a same relative clock, which is the starting point for distilling more general information about the duration of the operations. The *Clock Refinement* step performs this simplification over the redundant clocks.

In particular, we define two relative clocks $c_a$ and $c_b$ to be redundant if they are both reset and checked on the same transitions. Note that by construction each relative clock is reset on exactly one transition and is checked with a guard condition on exactly one transition. Thus, if $reset(c)$ is the transition where the relative clock $c$ is reset, and $check(c)$ is the transition where the relative clock $c$ is checked with a guard condition, two clocks $c_a$ and $c_b$ are redundant if $reset(c_a) = reset(c_b)$ and $check(c_a) = check(c_b)$.

*TkT* simplifies redundant clocks by dropping one of the two clocks, namely $c_b$, and its reset operation, and renaming all the occurrences of the dropped clock in its guard condition with the redundant clock that survives, namely all occurrences of $c_b$ are replaced with $c_a$. This transformation is performed for all the redundant clocks producing an *Automaton with Refined Clocks*, as shown in Figure 1. For instance, clocks $c2$, $c4$, $c9$ are all renamed as clock $c2$. Note that the refined automaton uses fewer clocks than the *Initial Automaton*, with each clock being associated with multiple observations.

Clock refinement is not applied to the clock that measures the absolute time because it cannot be redundant with any other clock. Note that the state merging process is already sufficient to accumulate multiple observations on each transition for the clock measuring the absolute time, contrarily to the relative clocks that need this refinement step.

### F. Guards Generation

The *Guards Generation* step iterates on the transitions of the *Automaton with Refined Clocks* and applies a *guard generation policy* to the data available on each transition to produce the *guard* of the transition. The guard generation policy is a function that takes a set of equality constraints on a same clock as input and generates a guard on that same clock as output. When the data associated with a transition refer to multiple clocks, the policy is also applied multiple times, once on each subset of equality constraints on a same clock. For instance, when the set of equality constraints {`t`=30, `t`=23, `c6`=11, `c6`=7} associated with the transition labeled `processItem` in the *Automaton with Refined Clocks* shown in Figure 1 is processed, the guard generation policy is applied twice, once on the data about clock `t` (i.e., {`t`=30, `t`=23}) and once on the data about clock `c6` (i.e., {`c6`=11, `c6`=7}).

We defined two concrete guard generation policies: the *min-max $\epsilon$-policy* and the *$\gamma$-confidence policy*. Given a set of observations for a clock `c`, the min-max $\epsilon$-policy generates a guard that constraints the value of `c` to the interval $[(1-\epsilon)min, (1+\epsilon)max]$, where $min$ and $max$ are the minimum and maximum values of `c` in the input values, and $\epsilon$ is a value in the range $[0, 1]$. Small values of $\epsilon$ bound the interval to the range of values provided as input, while large values of $\epsilon$ produces a more flexible interval, up to $[0, 2max]$ in case $\epsilon = 1$. Figure 1 shows the *Timed Automaton* generated with the min-max 0-policy.

The *$\gamma$-confidence policy* generates an interval that has a cumulative probability being equal to $\gamma$ to include the possible duration of an operation assuming a normal distribution of the durations. We consider $\gamma = 0.95$ and $\gamma = 0.99$ as possible values of the parameter. If the interval generated by the *$\gamma$-confidence policy* does not include all the values observed for a clock, *TkT* extends the interval until including the min/max values of the clock.

If only one value has been observed for a clock, both policies do not generate any guard.

Note that *TkT* generates timed automata that accept every timed trace used for the inference by construction. In fact, the *Initial Automaton* accepts exactly the timed traces provided as input. The state-merging process may only increase the combination of events that might be accepted by the model and cannot drop any behavior. The *Clock Refinement* step simplifies the model without altering the set of accepted behaviors. Finally, the *Guards Generation* step can only increase the range of acceptable timing for the events in the traces.

## III. Empirical Evaluation

To empirically evaluate *TkT*, we investigated the following five research questions.

**RQ1: Is *TkT* able to infer generic models that *comprehensively* capture the behavior of the software?** RQ1 investigates the sensitivity of *TkT*, that is its capability to derive models that accept legal traces, including the ones that have not been used for the inference, compared to Perfume.

**RQ2: Is *TkT* able to infer precise models that *reject* anomalous software behaviors?** RQ2 investigates the specificity of *TkT*, that is its capability to derive models that reject anomalous traces, for example traces recorded while the system is overloaded, compared to Perfume.

**RQ3: How does *TkT balance* the abilities to deal with legal and illegal behaviors?** RQ3 investigates how *TkT* balances sensitivity and specificity, compared to Perfurme.

**RQ4: How does *absolute time* affect the effectiveness of *TkT*?** RQ4 investigates how the generation of a clock that measures the absolute time affects sensitivity and specificity.

**RQ5: How does *TkT scale* with the size of the traces?** RQ5 investigates how the performance of *TkT* scales with the number of events that are processed, compared to Perfume.

### A. Prototype and Experiment Setup

The *TkT* prototype that we used for the experiments is implemented in Java and is available at http://www.lta.disco.

unimib.it/tools/tktail. Our implementation supports all the configurations and parameters described in this paper.

There are two main parameters that may influence how the timing behavior is encoded in the inferred model. The first parameter controls if guards on the absolute clock are generated. The second parameter controls the specific guard generation policy that *TkT* must use to generate guards, among the *min-max $\epsilon$-policy* and the *$\gamma$-confidence policy*. Both policies can be customized according to a parameter that can be assigned with real values in the range $[0..1]$.

In our empirical evaluation, we study both guard generation policies covering a range of values for their parameters. In the min-max $\epsilon$-policy, we cover the whole domain, sampling more densely the domain for small values of $\epsilon$, where we expect the technique to be more sensitive to changes. In the $\gamma$-confidence policy, we consider $\gamma$ equals to 0.95 and 0.99, to focus on guards that may cover the possible durations with high confidence. These configurations are investigated both with and without the generation of the guards on the absolute clock. Table I summarizes the set of configurations considered in our evaluation. Each configuration has an identifier.

The behavior of *TkT* is also influenced by the choice of parameter *k*, which does not play a role in guards generation but affects the identification of the states to be merged. In our empirical evaluation, we focus on the parameters that may influence the timing behavior, which is the novel aspect introduced in *TkT* compared to k-Tail, and we do not study the impact of *k* in the state merging process which has been already studied in many other papers [6], [13], [14], [26], [27], [28]. We rather take advantage of these results to fix the value of *k* to 2, which has demonstrated to be a good choice when analyzing traces collected from software systems.

TABLE I
*TkT* CONFIGURATIONS

| ID | | Policy | $\epsilon$ | $\gamma$ |
|---|---|---|---|---|
| abs clock | no abs clock | | | |
| M1 | M2 | min-max $\epsilon$-policy | 0.05 | - |
| M3 | M4 | min-max $\epsilon$-policy | 0.10 | - |
| M5 | M6 | min-max $\epsilon$-policy | 0.15 | - |
| M7 | M8 | min-max $\epsilon$-policy | 0.20 | - |
| M9 | M10 | min-max $\epsilon$-policy | 0.25 | - |
| M11 | M12 | min-max $\epsilon$-policy | 0.50 | - |
| M13 | M14 | min-max $\epsilon$-policy | 0.75 | - |
| M15 | M16 | min-max $\epsilon$-policy | 1.00 | - |
| G1 | G2 | $\gamma$-confidence policy | - | 0.95 |
| G3 | G4 | $\gamma$-confidence policy | - | 0.99 |

## B. Subjects of the Study

The study has been conducted on two sets of Java programs, each set generating invalid traces for different reasons. In particular, the first set of programs consists of the implementation of four well-known *algorithms*: the *merge sort* sorting algorithm [29], the *Rabin Karp* pattern matching algorithm [30], the *LZW* compression algorithm [31], and the *LZWdecom* decompression algorithm [31]. We used these algorithms to evaluate the capability of *TkT* and Perfurme to discriminate the valid and the invalid executions caused by *anomalies in the execution environment* (e.g., due to overloaded resources). To study this capability, we logged the execution of all the methods in the classes that implement the algorithms both when the system is not overloaded and when it is overloaded by four other processes intensively executing I/O operations. For each algorithm we produced 100 valid and 100 invalid traces.

The second set consists of five versions of the Guava *library* [32] affected by different performance problems. For our empirical study we selected the five most recent performance faults in the Guava issue tracker [33] (excluding performance enhancements) whose identifiers are: 371, 1013, 1155, 1196, and 1197. We used these library versions to evaluate the capability of *TkT* and Perfume to discriminate the valid executions from the invalid executions caused by *performance faults*. To study this capability, for each version of the library we generated the traces by running the original test suite distributed with Guava and a non-regression test suite that we implemented to extensively sample the fixed functionality with various inputs. We obtained the valid traces by running the test cases on the fixed version of the program (we extracted the fixes from the version history of Guava). We obtained the invalid traces by executing the same test cases on the faulty version of the software.

Since not every test execution of the faulty program necessarily produces a performance failure, we used the overhead observed when running the test suite of the program to objectively discriminate between valid and invalid traces (i.e., failed executions). In particular, the executions showing an overhead smaller than the one observed while running the test suite of the program have been classified as valid traces, consistently with Guava developers who have not recognized the performance faults in the library when running the Guava test suite. While the executions showing a higher overhead have been classified as invalid traces. Depending on the library version, this strategy allowed to generate between 340 and 1000 valid traces and between 51 and 501 invalid traces caused by actual performance faults.

In the evaluation, we recorded the traces using an AspectJ advice [34] that intercepts both method entry and exit events.

## C. *RQ1: Is* TkT *able to infer generic models that* comprehensively *capture the behavior of the software?*

*Procedure:* Research question RQ1 investigates the *sensitivity* of *TkT*, that is its capability to infer models that can accept the legal traces not occurring in the set of traces used for the inference. Since the quality of the inferred models depends not only on the algorithm, but also on its configuration and the completeness of the traces used for the inference, we studied the effectiveness of *TkT* for all the 20 configurations listed in Table I and for sets of traces of various sizes. We performed the study for both the four algorithms and the five versions of the Guava library.

In order to compare *TkT* to Perfume, we also executed Perfume against the traces collected from the four algorithms.

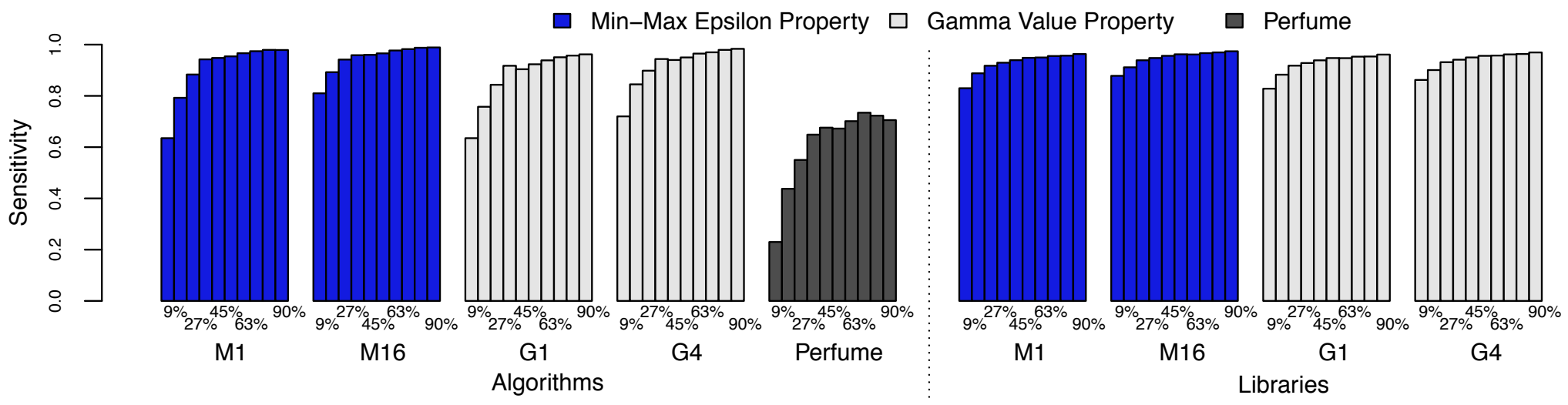

Fig. 2. RQ1: Sensitivity wrt Available Traces included in the Training Set.

We do not have results about Perfume applied to the traces collected from the Guava library because Perfume has not been able to infer a model after 10 hours of computation, which was our time limit for each experiment, in contrast to *TkT* that completed every inference task with less than 5 minutes (see RQ5 for a discussion about scalability).

To measure the impact of the number of traces on the sensitivity of the algorithms, we executed *TkT* and Perfume on random subsets of traces, considering from 10% to 100% of the set of valid traces. To mitigate the effect of randomness, we repeated the extraction process 10 times and reported average values. Note that these subsets of traces may miss many valid application behaviours facilitating the generation of incomplete models.

For each set of traces, we used the 10-fold cross validation method to measure the sensitivity of *TkT* and Perfume. The 10-fold cross validation method works by dividing an initial set of traces in 10 sets of the same size (folds), using 9 sets for the inference the model (training set) and 1 set for the validation of the model (validation set). The sensitivity of the algorithm is computed as the fraction of the traces in the validation set accepted by the model inferred from the training set. The process is repeated 10 times, every time using a different fold as validation set. The final score is obtained as the average value of the sensitivity for the 10 folds. This entire process is repeated 5 times to mitigate the effect that randomly partitioning a set of traces into 10 folds may have. Overall, the study on the sensitivity required the generation of 21 models (20 for *TkT* and 1 for Perfume) from each training set, for each subset of traces, for each subject application (except for library cases, where Perfume did not complete), and for 5 repetitions, for a total of 92,000 models inferred.

*Results:* The barplots in Figure 2 show the sensitivity of the models inferred using *TkT* and Perfume when an increasing number of traces is available for the inference. In particular, each barplot represents a different configuration and each vertical bar shows the average fraction of accepted traces among the subject applications.

For both the min-max $\epsilon$-policy and $\gamma$-confidence policy we report the configurations with the lowest (*M1* and *G1*) and the highest sensitivity values (*M16* and *G4*). We also report the sensitivity of Perfume.

The results show that the sensitivity of *TkT* is generally high for both the algorithms and the five versions of the Guava library. Both policies perform well, with the min-max $\epsilon$-policy performing slightly better than the $\gamma$-confidence policy in the case of the algorithms. This is probably due to the availability of a good number of traces that broadly covers the valid behaviors, which can be better exploited by the min-max $\epsilon$-policy.

Finally, *TkT* outperforms Perfume in terms of sensitivity. Perfume rejects many traces due to the presence of constraints that overfit the input samples used for the inference. For instance, Perfume never reaches a score of 0.8 for the sensitivity even when using 90% of the available traces for the inference, while the min-max $\epsilon$-policy already achieves higher values of sensitivity when 18% of the available traces is used for the inference.

## *D. RQ2: Is* TkT *able to infer precise models that* reject *anomalous software behaviors?*

*Procedure:* Research question RQ2 investigates the *specificity* of *TkT*, that is its capability to reject invalid traces. In our experiments invalid traces are characterized by the presence of anomalous timing due to either system overloading or performance faults.

We measure the specificity of a model as the fraction of invalid traces rejected by the model. We studied the specificity for the same range of configurations and size of the training sets used for RQ1.

*Results:* The barplots in Figure 3 show the specificity of the models inferred with *TkT* and Perfume when an increasing number of traces is available for the inference. In particular, each barplot represents a different configuration and each vertical bar shows the average fraction of rejected traces among the four subject applications.

For both the min-max $\epsilon$-policy and $\gamma$-confidence policy we report the configurations with the highest (*M1* and *G1*) and the lowest (*M16* and *G4*) specificity values. We also report the specificity of Perfume limitedly to the case of the overloaded environment because it has not been able to infer the models for the Guava library.

Perfume achieves a perfect specificity for any number of input traces. This is due to the overfitting of the timing constraints generated by Perfume, which can easily reject any trace that includes a small variation of the timing behavior observed on the traces used for the inference.

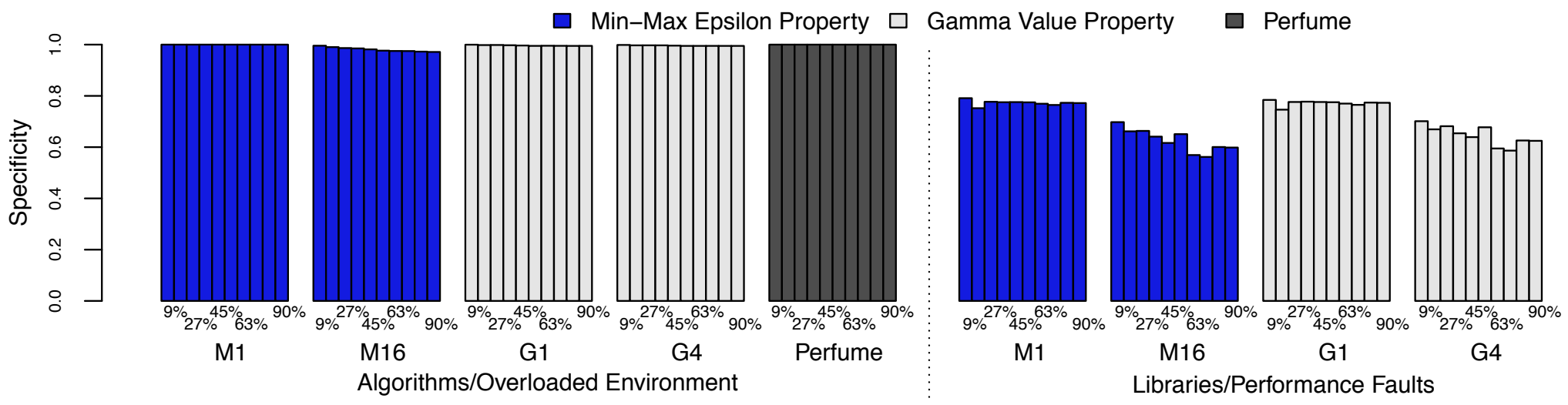

Fig. 3. RQ2: Specificity wrt Available Traces included in the Training Set.

For all the configurations and cases *TkT* has been able to identify more than 60% of the invalid traces on average. *TkT* achieves almost the same result of Perfume when applied to the invalid traces obtained from an overloaded environment. The specificity is lower when applied to performance faults. This result is due to a single performance fault (fault number 1196) out of the five that have been studied. This fault produces traces with an extremely small overhead, which is sufficient to classify traces as invalid according to our criterion, but that can be hardly recognized by *TkT*, and could also be hardly recognized by a developer. It is worth to mention that the average specificity of *TkT* for the other four performance faults is 0.86.

These results indicate that the models inferred with *TkT* can be very effective in discriminating both classes of invalid executions (overloaded environment and performance faults), as long as the overhead is not so marginal to be hardly recognizable by a user.

### E. *RQ3: How does* TkT *balance the abilities to deal with legal and illegal behaviors?*

*Procedure:* RQ3 investigates how *TkT* balances sensitivity and specificity, compared to Perfume. Finding a balance between sensitivity and specificity is important to prevent the generation of models that may trivially reject or accept all the traces. In particular, a good model should tolerate small changes on the timing behavior, promptly reporting any significant deviation.

To study this balance, we computed the harmonic mean of sensitivity and specificity for all the configurations that we studied in RQ1 and RQ2. We aggregated data according to the percentage of available traces used for the inference. We distinguished between a small portion of traces (at most 30% of the available traces used for the inference), an intermediate portion of traces (the number of traces used for inference comprised between 30% and 70%), and a high portion (more than 70%) of the available traces.

*Results:* The barplots in Figure 4 show the results for all the configurations when a small, intermediate, and high number of traces are available.

*TkT* performed better than Perfume, with larger differences when fewer traces are available. The various *TkT* policies performed similarly. The min-max $\epsilon$-policy performs better than the $\gamma$-confidence policy on the algorithms (this is the direct consequence of the higher sensitivity already discussed in RQ1). On the library versions the two policies perform similarly, with configurations using absolute clocks, which correspond to the bars at an odd position, achieving better results (this is a consequence of the higher specificity obtained by these configurations, which is further discussed in RQ4).

The plots also show that *TkT* obtains better results than Perfume (see the bar named *Pf*) independently on the percentage of traces used for the inference.

### F. *RQ4: How does* absolute time *affect the effectiveness of* TkT*?*

*Procedure:* RQ4 investigates how the use of the absolute clocks affects sensitivity and specificity. The adoption of absolute time may allow to detect operations that begin or terminate too late with respect to the beginning of the trace, but may also introduce subtle dependencies among the operations. To investigate the effect of adopting absolute time, we compare the sensitivity and the specificity of the configurations that use absolute clocks to the configurations that do not use it, for all the available cases.

*Results:* The box plot in Figure 5 shows the results for sensitivity and specificity with and without absolute clocks (dotted lines show average values). According to our set of experiments the adoption of absolute clocks leads to a small decrease of sensitivity while they contribute to a higher specificity. Please note that the high variance in specificity results mainly depends on the performance fault discussed in Section III-D where *TkT* has shown a low specificity.

We conjecture that the negative effect of absolute clocks might be stronger when the application under test includes several cyclic behavior repeated for very long executions. Further experiments on specific domains, such as software systems running for long periods like application servers, might be useful to additionally study benefits and drawbacks of absolute clocks.

### G. *RQ5: How does* TkT *scale with the size of the traces?*

*Procedure:* To investigate RQ5, we compare the inference time of *TkT* to the inference time of Perfume. In particular, we investigate how the cost of the inference increases when an increasing number of traces are processed by the algorithms. We measured the time necessary to infer a model for all the executions performed to answer RQ1. All the experiments have been run on a Lenovo System X 3500 M5 Server with 32 cores (3.20GHz). Since Perfume did not complete on the cases based on Guava, we only considered the four algorithms for this study.

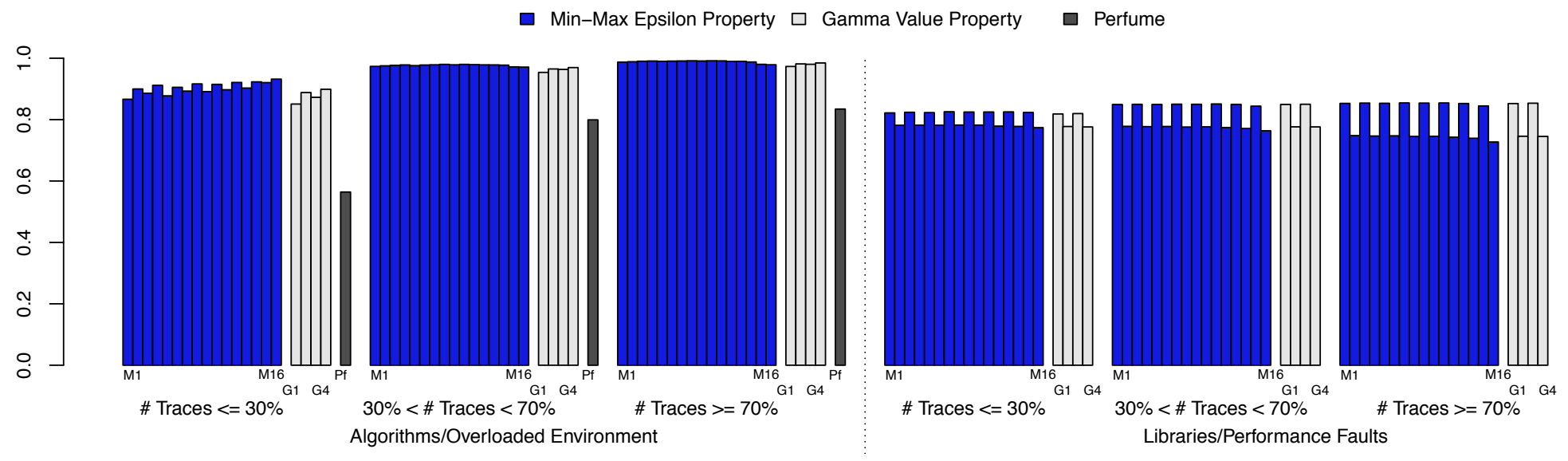

Fig. 4. RQ3: Harmonic Mean of Specificity and Sensitivity wrt Available Traces included in the Training Set.

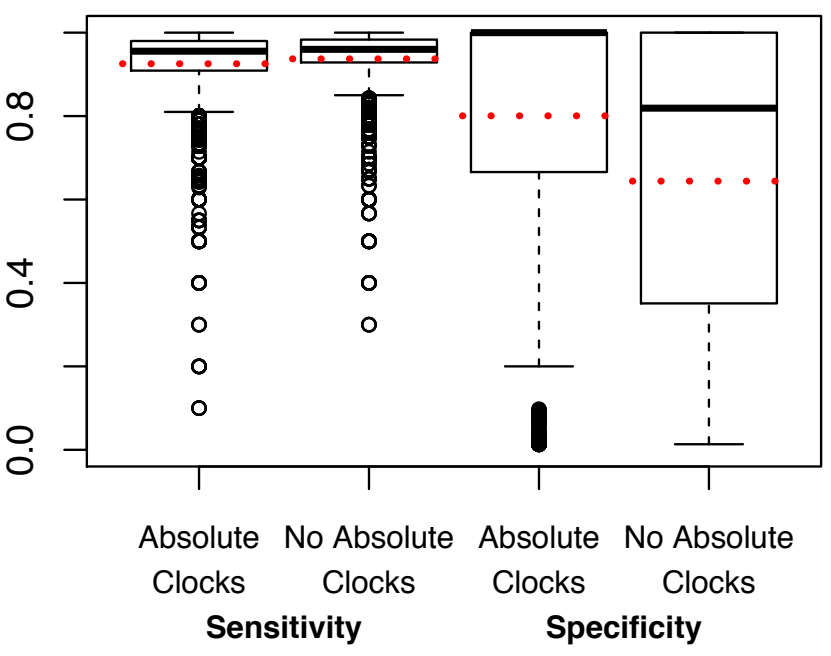

Fig. 5. RQ4: Distribution of sensitivity/specificity with and without absolute clocks. Dotted lines show the average across all the inferred models.

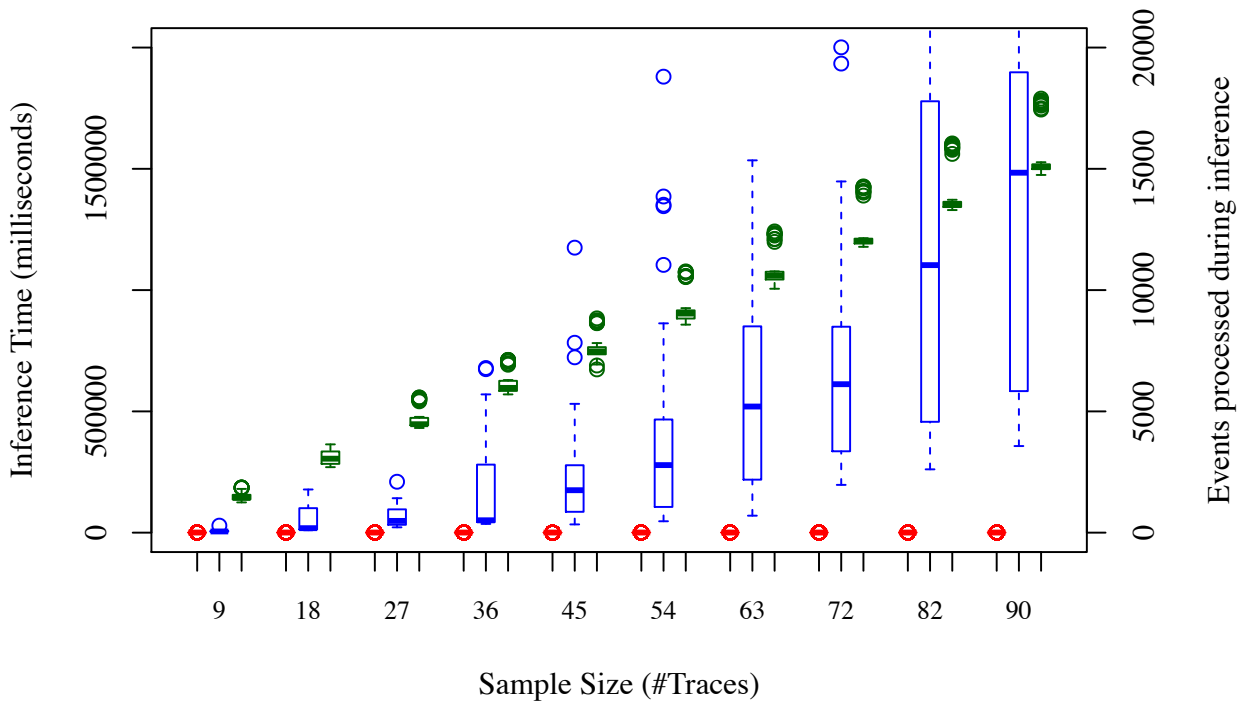

Fig. 6. RQ5: Inference time of *TkT* (red) and Perfume (blue), plus number of events processed during inference (green).

*Results:* Figure 6 shows a boxplot with the inference time of *TkT* compared to Perfume. Each box represents multiple executions all performed with samples of the same size.

Data show that Perfume is significantly slower than *TkT*. For example when the number of traces to be processed is close to 100, *TkT* is almost three order of magnitude faster than Perfume (Perfume takes on average 34.7 minutes, while *TkT* takes 0.5 seconds). More specifically the cost of running Perfume increases linearly with respect to the number of processed events, while the cost of running *TkT* growths by a negligible fraction.

These results suggest that Perfume might hardly scale to large systems producing long traces, as we experienced with the five Guava versions that produced more than 20k events to be processed, contrarily to *TkT* that scaled gracefully with the number of events.

## IV. Related Work

Specification mining includes a variety of techniques and approaches for generating models that represent the behavior of software systems from sets of execution traces. Here we discuss specification mining techniques according to the type of model that they generate: ordering of operations, simple FSAs, annotated FSAs, and FSAs with time information.

*Operations Order:* Many techniques can generate models about the ordering of operations. Some approaches focus on partial models, that is models that constraint the ordering of a portion of the operations that can be executed by a system. A popular way to express properties about partial order of operations is with temporal rules that specify temporal relations among events, without specifying the full behavior of the software. Various classes of temporal rules can be mined with techniques such as Perracotta [16], the algorithm by Lo et al. [35], and Texada [17].

In contrast with these approaches, which are useful when the developer is interested in properties about a subset of the events that can be produced by a software system, *TkT* aims to capture the whole behavior of a software system with a finite state model that represents every possible combination of events that can be produced by the monitored software.

*Mining Simple FSAs:* Mining accurate finite state models from execution traces is a long living problem. In their seminal work, Biermann and Feldman introduced k-Tail [25], a well-known algorithm for the generation of a finite state automaton from a set of observations using an algorithm based on iterative state merging. The k-Tail algorithm has inspired many other techniques that modified and improved the learning process in different ways, for instance making the resulting model more compact [26] and introducing a steering process to improve the quality of the model [13]. Also *TkT* uses a modified version of the state merging process originally introduced in k-Tail.

When the traces include information about the state of the monitored application, it is possible to apply a style of inference that exploits state information rather than the sequences of the events. Some of the approaches that apply this strategy are ADABU [36], ReAjax [37], and Revolution [38]. These techniques can be extremely effective, but they can be applied only when suitable monitors that can efficiently inspect the state of the application can be implemented. We designed *TkT* to be applicable to traces that are frequently available and easy to produce, such as traces with events and timestamps, for this reason *TkT* does not rely on the existence of monitors that can extract additional state information.

*Mining Annotated FSAs:* Few specification mining techniques deal with the problem of deriving FSAs that include additional infromation necessary to better capture the behavior of an application. This is the case of specification mining techniques that infer extended finite state machines [14], [15] (i.e., FSAs with transitions annotated with guards conditions specifing the values that can be assigned to some variables), and FSA annotated with data-flow information [4] (i.e., FSAs where transitions are labeled with identifiers that can capture how the values of some variables reoccur across events). While the annotated models can represent behavioral information that cannot be captured with simple FSAs, inferring extended models may negatively affect the quality of the models in some of the cases [27].

*TkT* also mines behaviors that cannot be captured with simple FSAs. Contrarily to existing extensions that focus on adding different classes of constraints on the models, *TkT* focuses on time information, which is an aspect that requires a specific extension to the mining process. The time information represented in the model produced by *TkT* requires the generation of clocks, resets, and guards on clocks, which cannot be represented with the existing extended models.

*Mining FSAs with Time Information:* Mining models that represent both the functional behavior of an application, such as the operations that can be performed and their ordering, and time information, such as the time required by each operation to complete, is important to be able to work on the interplay between these two aspects.

Perfume [5] is a recent technique that addresses the problem of mining models that capture the interplay between the functional and the timing behavior by extending Synoptic [10] with the ability to deal with the durations of the operations.

In addition to Perfume, several techniques for the inference of automata with time information [18], [19], [20], [21], [22] have been developed in recents years, but most of them have been specifically designed to model embedded and cyber-physical systems. OTALA [18] derives timed automata where each distinct combination of discrete input/output signals corresponds to a state of the automata, a criterion that cannot be easily applied to traces recorded from other software systems. BUTLA [19] instead, derives a prefix tree acceptor that captures the sequences of system events appearing in traces. BUTLA integrates a state merging criterion that merges states with a similar probability of being final, or receiving the same events. This merging criterion is effective in the context of cyber-physical systems where the system state may directly depend on individual events (e.g., a switch turned on), but it is hardly applicable to other types of software systems where the current system state may depend on sequences of operations. HyBUTLA [20] augments BUTLA with the capability of dealing with discrete and continuos signals. Contrarily to BUTLA and HyBUTLA, *TkT* targets generic software systems that can execute sequences of possibly nested operations by suitably extending the state merging strategy originally defined in kTail.

Several approaches [18], [20], [21], [22] derive *real-time automata* [21], that is one-clock timed automata [24] where constraints simply capture the time difference between consecutive events. *TkT* instead derives constraints that can bound the duration of pair of non-consecutive events (e.g., the beginning and the end of the execution of a method), thus being able to effectively deal with the duration of nested operations. This same limitation characterizes also approaches that derive event recording automata [23].

## V. Conclusions

Mining accurate models that can serve software engineering tasks is challenging. So far specification mining solutions focused on the generation of behavioral models that capture the functional behavior of a system, producing models like temporal rules [16], [35], [17], finite state machines [25], [37], [13], and extended finite state machines [14], [15].

Although these models are useful, they fail to capture the interplay between the functional and the timing behavior of a system. This relation is relevant in many contexts where completing operations with unusual timing might represent a problem. For instance, an algorithm that requires too long to complete may cause serious inefficiencies in a system.

In this paper we presented *TkT*, a specification mining technique that can generate a timed automaton from a set of execution traces. Timed automata can suitably represent both the functional and the timing aspects, supporting analyses that consider those two aspects independently or jointly. Our empirical evaluation provided evidence of the effectiveness and the efficiency of the solution.

Our future work includes applying TkT to a larger body of systems, and further improving its effectiveness, for example by introducing mechanisms to detect the faults that can be hardly captured right now, such as faults that produce a very small, but unjustified, overhead.

## Acknowledgment

This work has been partially supported by the H2020 Learn project, which has been funded under the ERC Consolidator Grant 2014 program (ERC Grant Agreement n. 646867).

# References

[1] V. Dallmeier, N. Knopp, C. Mallon, G. Fraser, S. Hack, and A. Zeller, "Automatically generating test cases for specification mining," *IEEE Transactions on Software Engineering (TSE)*, vol. 38, no. 2, pp. 243–257, 2012.

[2] C. Pacheco and M. D. Ernst, "Eclat: Automatic generation and classification of test inputs," in *Proceedings of the European Conference on Object-Oriented Programming (ECOOP)*, 2005.

[3] O. Raz, P. Koopman, and M. Shaw, "Semantic anomaly detection in online data sources," in *Proceedings of the International Conference on Software Engineering (ICSE)*, 2002.

[4] L. Mariani and F. Pastore, "Automated identification of failure causes in system logs," in *Proceedings of the International Symposium on Software Reliability Engineering (ISSRE)*, 2008.

[5] T. Ohmann, M. Herzberg, S. Fiss, A. Halbert, M. Palyart, I. Beschastnikh, and Y. Brun, "Behavioral resource-aware model inference," in *Proceedings of the International Conference on Automated Software Engineering (ASE)*, 2014.

[6] L. Mariani, F. Pastore, and M. Pezzè, "Dynamic analysis for diagnosing integration faults," *IEEE Transactions on Software Engineering (TSE)*, vol. 37, no. 4, pp. 486–508, 2011.

[7] S. Hangal and M. S. Lam, "Tracking down software bugs using automatic anomaly detection," in *Proceedings of the International Conference on Software Engineering (ICSE)*, 2002.

[8] A. Babenko, L. Mariani, and F. Pastore, "Ava: Automated interpretation of dynamically detected anomalies," in *proceedings of the International Symposium on Software Testing and Analysis (ISSTA)*, 2009.

[9] M. Gabel and Z. Su, "Testing mined specifications," in *Proceedings of the ACM SIGSOFT International Symposium on the Foundations of Software Engineering (FSE)*, 2012.

[10] I. Beschastnikh, Y. Brun, S. Schneider, M. Sloan, and M. D. Ernst, "Leveraging existing instrumentation to automatically infer invariant-constrained models," in *Proceedings of the Joint Meeting on European Software Engineering Conference and Foundations of Software Engineering (ESEC/FSE)*, 2011.

[11] I. Beschastnikh, Y. Brun, J. Abrahamson, M. D. Ernst, and A. Krishnamurthy, "Using declarative specification to improve the understanding, extensibility, and comparison of model-inference algorithms," *IEEE Transactions on Software Engineering (TSE)*, vol. 41, no. 4, pp. 408–428, 2015.

[12] M. D. Ernst, J. Cockrell, W. G. Griswold, and D. Notkin, "Dynamically discovering likely program invariants to support program evolution," *IEEE Transactions on Software Engineering (TSE)*, vol. 27, no. 2, pp. 99–123, 2001.

[13] D. Lo, L. Mariani, and M. Pezzè, "Automatic steering of behavioral model inference," in *Proceedings of the Joint Meeting on European Software Engineering Conference and Foundations of Software Engineering (ESEC/FSE)*, 2009.

[14] D. Lorenzoli, L. Mariani, and M. Pezzè, "Automatic generation of software behavioral models," in *Proceedings of the International Conference on Software Engineering (ICSE)*, 2008.

[15] N. Walkinshaw, R. Taylor, and J. Derrick, "Inferring extended finite state machine models from software executions," *Journal of Empirical Software Engineering (EMSE)*, pp. 1–43, 2015.

[16] J. Yang, D. Evans, D. Bhardwaj, T. Bhat, and M. Das, "Perracotta: mining temporal API rules from imperfect traces," in *Proceeding of the International Conference on Software Engineering (ICSE)*, 2006.

[17] C. Lemieux, D. Park, and I. Beschastnikh, "General LTL specification mining," in *Proceedings of the International Conference on Automated Software Engineering (ASE)*, 2015.

[18] A. Maier, "Online passive learning of timed automata for cyber-physical production systems," in *2014 12th IEEE International Conference on Industrial Informatics (INDIN)*, July 2014, pp. 60–66.

[19] A. Maier, A. Vodenčarevič, O. Niggemann, R. Just, and M. Jäger, "Anomaly detection in production plants using timed automata," in *8th International Conference on Informatics in Control Automation and Robotics (ICINCO)*, 2011, pp. 363–369.

[20] O. Niggemann, B. Stein, A. Vodencarevic, A. Maier, and H. K. Bning, "Learning behavior models for hybrid timed systems," in *Proceeding of the 26th AAAI Conference on Artificial Intelligence (AAAI'12)*, 2012, pp. 1083–1090.

[21] S. Verwer, M. De Weerdt, and C. Witteveen, "An algorithm for learning real-time automata," in *Benelearn 2007: Proceedings of the Annual Machine Learning Conference of Belgium and the Netherlands*, 2007.

[22] J. Schmidt and s. Kramer, "Online induction of probabilistic real time automata," in *Proceedings of the International Conference on Data Mining (ICDM)*, 2012.

[23] O. Grinchtein, B. Jonsson, and P. Pettersson, "Inference of event-recording automata using timed decision trees," in *Proceedings of the International Conference on Concurrency Theory (CONCUR)*, 2006.

[24] C. Martin-Vide, B. Truthe, S. Verwer, M. de Weerdt, and C. Witteveen, "The efficiency of identifying timed automata and the power of clocks," *Information and Computation*, vol. 209, no. 3, pp. 606 – 625, 2011.

[25] A. Biermann and J. Feldman, "On the synthesis of finite-state machines from samples of their behavior," *IEEE Transactions on Computers*, vol. 21, no. 6, pp. 592 – 597, 1972.

[26] J. Cook and A. Wolf, "Discovering models of software processes from event-based data," *ACM Transactions on Software Engineering and Methodology (TOSEM)*, vol. 7, pp. 215–249, 1998.

[27] D. Lo, L. Mariani, and M. Santoro, "Learning extended fsa from software: An empirical assessment," *Journal of Systems and Software (JSS)*, vol. 85, no. 9, pp. 2063 – 2076, 2012.

[28] S. P. Reiss and M. Renieris, "Encoding program executions," in *Proceedings of the International Conference on Software Engineering (ICSE)*, 2001.

[29] D. E. Knuth, *The Art of Computer Programming, Volume 1 (3rd Ed.): Fundamental Algorithms*. Redwood City, CA, USA: Addison Wesley Longman Publishing Co., Inc., 1997.

[30] R. M. Karp and M. O. Rabin, "Efficient randomized pattern-matching algorithms," *IBM Journal of Research and Development*, vol. 31, no. 2, pp. 249–260, 1987.

[31] T. A. Welch, "A technique for high-performance data compression," *Computer*, vol. 17, no. 6, pp. 8–19, June 1984.

[32] Google, "Java collections library," https://github.com/google/guava.

[33] ——, "Guava bug repository," https://github.com/google/guava/issues/.

[34] Eclipse, "Aspectj," https://eclipse.org/aspectj/, 2016.

[35] D. Lo, G. Ramalingam, V. P. Ranganath, and K. Vaswani, "Mining quantified temporal rules: Formalism, algorithms, and evaluation," *Science of Computer Programming*, vol. 77, no. 6, pp. 743–759, 2012.

[36] V. Dallmeier, C. Lindig, A. Wasylkowski, and A. Zeller, "Mining object behavior with ADABU," in *Proceedings of the International Workshop on Dynamic Analysis (WODA)*, 2006.

[37] A. Marchetto, P. Tonella, and F. Ricca, "State-based testing of Ajax web applications," in *Proceedings of the International Conference on Software Testing, Verification, and Validation (ICST)*, 2008.

[38] L. Mariani, A. Marchetto, C. Nguyen, P. Tonella, and A. Baars, "Revolution: Automatic evolution of mined specifications," in *Proceedings of the International Symposium on Software Reliability Engineering (ISSRE)*, 2012.